# Contact-induced molecular reorganization in E. coli model lipid membranes

Nicolo Tormena[1], Teuta Pilizota[2,3,*], Kislon Voitchovsky[1,*]

1. Physics Department, Durham University, South Road, Durham DH1 3LE, UK
2. School of Biological Sciences and Centre for Engineering Biology, The University of Edinburgh, Alexander Crum Brown Road, Edinburgh, EH9 3FF, UK
3. Department of Physics, University of Cambridge, JJ Thomson Avenue, Cambridge CB3 0HE

*Teuta Pilizota and Kislon Voitchovsky

**Email:** tp579@cam.ac.uk, kislon.voitchovsky@durham.ac.uk

**Abstract**

Biological membranes are complex, dynamic structures essential for cellular compartmentalization, signaling, and mechanical integrity. The molecular organization of eukaryotic membranes has been extensively studied, including the lipid raft-mediated lateral organization and the influence of the specific molecular interactions. Bacterial membranes were traditionally viewed as compositionally simpler and structurally uniform. Recent evidence, however, reveals that they possess significant lipid diversity and can form functional microdomains reminiscent of eukaryotic lipid rafts, despite lacking sterols and sphingolipids. Yet, the impact of unspecific physical contacts on the local molecular organization and evolution of the prokaryotic membranes remains poorly understood. Here we use a model lipid membrane mimicking the composition of Escherichia coli's inner membrane to investigate the impact of contacting substrates on the membrane nanoscale evolution, when close to its transition temperature, $T_m$. As expected, the presence of a substrate lowers the $T_m$ by ~10 °C and induces a differential leaflet transition. However, it also slows down the phase transition kinetics by almost 2 orders of magnitude while simultaneously enabling a spinodal-like lateral molecular reorganization. This creates local alterations of the phase of the membrane, with the emergence of mechanically stiffer, yet still fluid nanodomains evolving over substrate-dependent timescales, consistent with a substrate-biased lipid flip-flop mechanism. The results verify previous theoretical predictions and demonstrate that a general physical mechanism—driven by membrane-surface interactions—can spontaneously induce lipid domain formation in bacterial membranes. This is bound to have notable consequences for its function and mechanical role, including in processes like osmotic pressure regulation.

**Significance Statement**

Both eukaryotic and prokaryotic membranes typically exist close to their transition temperature and actively readjust their molecular composition and organization depending on external stimuli and functional needs. Here we show that physical contact between a model bacterial membrane and the surface of solids triggers a local rearrangement of the membrane with significant changes in the mechanical properties and phase behavior—without the need for specialized lipid components. These findings challenge long-standing assumptions about the simplicity of prokaryotic membranes and highlight a fundamental, passive mechanism for membrane organization. Furthermore, the findings have important implications for understanding bacterial physiology and the role membranes play in regulating cellular processes, like osmotic stress adaptation.

**Main Text**

**Introduction**

Biological membranes separate different cellular compartments and the cells themselves from the external environment, and comprise lipids organized in a bilayer structure together with other amphiphilic molecules, and embedding a very large number of proteins and sugars (1). They are essential for achieving and controlling cellular homeostasis, signaling and uptake of macromolecules and ions, as well as bearing mechanical load (1–4). Extensive research has focused on the relationship between the compositional complexity of biological membranes and the resulting molecular organization and structure across scales (5–10). Early studies on model membrane systems composed solely of lipids revealed distinct physical phases that depend on structural properties of the lipids (11, 12). These range from tightly packed solid-ordered ($S_o$) to more fluid, disordered phases termed liquid disordered ($L_d$). Subsequent studies on plasma membranes of eukaryotic cells indicated the existence of intermediate liquid-ordered ($L_o$) phase domains, where lipids are more tightly packed than in the $L_d$ phase but still retain a degree of lateral mobility (13, 14). Because such $L_o$ domains, often termed lipid rafts, are small and temporally dynamic structures, their existence has been extensively debated over the last decade (15). The current consensus not only recognizes their existence *in vivo*, but also their role in regulating the uptake of extracellular vesicles (16), mechanosensing (17) and other signaling processes (8, 18), including hosting and regulating crucial membrane receptors for pathogen recognition (19, 20). Lipid rafts have a common lipidomic profile consisting of phospholipids or sphingolipids mixed with sterols such as cholesterol (21). Cholesterol, in particular, has been extensively studied for its role in controlling the fluidity of lipid rafts and the characteristics of the $L_o$ phase (22–24). With cholesterol only present in eukaryotic cells, the emergence of lipid rafts has often been viewed as a critical evolutionary step (25).

In contrast, prokaryotic membranes lack both sphingolipids and sterols and were long thought to be structurally simple. However, recent studies have revealed that bacterial membranes exhibit a far more diverse lipid composition than previously assumed (26–29). Bacterial lipids are remarkably varied, a complexity that has often been overlooked due to the dominance of nutrient-rich culture studies in the literature (26). Under stress or nutrient-depleted conditions, bacteria display extensive lipid diversity and engage in homeoviscous adaptation, a process by which they modify their membrane composition to maintain fluidity across different environmental conditions (26). Furthermore, recent studies on bacterial membranes provide strong evidence for functional, lipid raft-like, microdomains (30–34). Their formation is driven by lipid-lipid interactions, particularly in the presence of cardiolipin (30, 35, 36), carotenoids (37, 38), and hopanoids (39–41), which behave similarly to cholesterol and cluster within native lipid domains, also modulating the local membrane fluidity (30, 36, 42). The implication is that the $L_o$ domains in biological membranes predates the evolution of sterols (41).

Beyond lipid-lipid interactions, the biological membrane is not an isolated free-standing layer, but rather exists within a highly crowded cellular environment and is itself crowded with proteins (9, 43). These proteins, whether embedded within the bilayer or associated with external scaffolds and filamentous networks, interact with the bilayer and play a role in modulating the membrane's functional, mechanical, and topographical properties, as well as contributing significantly to their lateral organization (44–53). Even minimal physical contacts can significantly influence membrane domain behavior (54); points of contact are sufficient to trigger local phase transition or molecular reorganization in the membrane (53).

However, most of our understanding of this interplay comes from studies in eukaryotic systems, despite bacterial intracellular environment being even more crowded, and the recent evidence that *E. coli* membranes can bear mechanical load (3, 4). Furthermore, the bacterial cytoskeletal network (55, 56), which comprises homologues of eukaryotic actin and tubulin proteins, remains in close contact with the inner membrane throughout key cellular processes such as division (57, 58), proliferation (59), morphogenesis (60–62), and motility (63–65). Other membrane-proximal structures, such as the peptidoglycan layer (66–69) –the cell wall– are also in contact with membranes. The cell wall plays a shaping role and bears osmotic pressure, making it a target for many antibiotics.

While specific interactions between the membrane and contacting structures have been the focus of multiple studies (60, 62, 66, 70), the fundamental aspects of non-specific 'physical' contact on the local organization and evolution of bacterial membranes remain poorly understood. This gap is all the more significant considering the fact the main transition temperature of bacterial membrane is usually close to that of their environment (26): small perturbations are susceptible to induce large changes (53). Here we comparatively investigate the phase behavior and evolution of model *E. coli* inner lipid bilayer systems when supported on different substrates (supported lipid bilayers – SLBs) or suspended in solution. Consistent with previous studies (71–76), we identify a shift of ~10 °C in main transition temperature ($T_m$) between the free-standing vesicles and SLBs of the model membranes, with a leaflet-by-leaflet behavior. However, our results also reveal a striking dependence of the associated transition kinetics on the presence of a substrate even in the absence of any of the canonical $L_o$-promoting components such as cardiolipin, hopanoids, or sterols.

The stabilizing nature of lipid-substrate interactions has long been recognized (71–77), with the key aspect being the asymmetry it introduces between the membrane leaflets, simply because one is naturally closer to the support. This, in turn, has complex ramifications for the membrane's mechanical properties, including phase transition (78–82) and differential tension between the leaflets. At the molecular level, the interaction with a substrate can lead to altered diffusion in the contacting leaflet, molecular redistribution within leaflets (78, 79, 83), as well as the exchange of lipids between leaflets (78, 84–86), the so-called lipid flip-flop. Multiple studies have proposed specific molecular mechanisms underpinning these molecular reorganizations with the flip-flop stimulated by surface defects (84, 87, 88).

Here, we combine differential scanning calorimetry (DSC) and liquid atomic force microscopy (AFM) to track the slow structural rearrangements of the model inner *E. coli* membrane over >24 h. The observed molecular rearrangements are consistent with substrate-membrane interactions acting as flip-flop–promoting sites (80, 84, 87–89), locally biasing the lipids distribution between the two leaflets to create an asymmetric lipid composition. The existence of such a general mechanism for the spontaneous formation of nanodomains in prokaryotic membranes can have far-reaching consequences for our understanding of their organization and function, with the present work providing novel insights into their native morphology, including the recently discovered role they play in actively bearing osmotic pressure (3, 4).

**Results**

**Surface interactions directly impact the phase transition kinetics**

Before investigating the long-term evolution of our inner membrane *E. coli* model system, it is necessary to understand the influence of a static contacting solid (here a substrate) on the phase behavior of the membrane. We comparatively track the phase transition of unsupported bilayers in solution using DSC (90–93), and of planar SLBs using AFM. The unsupported bilayers serve as a reference; they are large vesicles (LVs) in aqueous solution and have an identical lipid composition to the SLBs (*see Materials and Methods*). DSC is a well-established technique for characterizing the thermodynamic properties of unsupported lipid systems, and AFM allows direct quantification of the fractions of $L_d$ and $S_o$ phases at each temperature over a given transition. In both cases, and in keeping with the stated goals of the study, the measurements are systematically taken upon cooling the sample down from a purely $L_d$ phase to a fully $S_o$ phase, with a range of temperature and cooling rates being explored.

The DSC results always exhibit a single peak regardless of the colling rate (Fig. 1A-B). The peak is characteristic of a standard first-order phase transition, with the cooling rate slightly influencing the transition temperature (maximum in Fig. 1A). This is a kinetic effect induced by the imposed temperature changes being too fast for the system to fully equilibrate across the transition. However, by comparing $T_m$ for both cooling and heating experiments, it is possible to infer an equilibrium transition temperature of 20.0 °C ± 0.1 °C (i.e. quasistatic cooling rate → 0°C/s) (94). Overall, the DSC results are in line with the results from previous calorimetric studies on similar vesicular systems (94, 95) and within the expected value range for lipid membranes (96).

The same transition is explored by AFM for SLBs, allowing direct visualization of the $L_d$ to $S_o$ phase transition as the lipids pack more tightly upon cooling (Fig. 1C-D) (73, 97). Unlike for the LVs in solution, two different $T_m$ values can be distinguished (see suppl. Fig. S1 for details). The interactions between the lipid headgroups and the solid substrate stabilize the lipids in the proximal leaflet and increase its $T_m$ relative to the distal leaflet. The effect, previously reported in the literature (73, 98), reflects the differences in local interactions and configurational entropy experienced by the lipids in the two leaflets. The existence of two distinct $T_m$ values spanning the significant range of temperatures seen in Figs 1C-D complicates direct comparison with the DSC results. Given our stated goal of using SLBs to ascertain the impact of the contacting solid, we hereafter focus only on the transition of the proximal leaflet, the first transition observed when cooling down the sample from the $L_d$ state. As expected, the SLB's proximal leaflet presents an almost 10°C higher $T_m$ than measured for the LVs (Fig. S2). The (near) equilibrium $T_m$ of 27.2 ± 0.6 °C for SLB's was estimated from 'quasi-static' cooling experiment with membrane allowed to equilibrate for 15 min between cooling steps of 2 °C (average cooling rate is $10^{-3}$ °C/min). The cooling is not continuous to favor relaxation of the membrane between cooling steps.
Having inferred the equilibrium $T_m$ values for LVs and SLBs, we turn our attention to the transition kinetics by repeating the experiments at different cooling rates. The results reveal a strong impact of the solid substrate (Fig. 2A) with the measured $T_m$ exhibiting an exponential dependence on the cooling rate (Fig. 2B, see also suppl. Fig. S2B-C). This result is not obvious considering the relatively weak interaction between lipid headgroups and the solid substrate underneath; several layers of water are present in-between with an overall water thickness ranging from 10 to 20 Å (99–101). For MLVs present in the LVs sample, interactions between lipid headgroups in adjacent bilayers could affect the transition kinetics, but the effect is negligible on our experimental timescale. Other thermodynamic parameters, such as the onset temperature and the enthalpy associated with the LVs and SLBs transitions do not show significant dependence on the cooling rate, as indicated in the summary graphs in suppl. Figs. S3, S4 and suppl. Table S1, S2.

**Surface interactions induce molecular reorganization in the SLBs**

To better understand substrate-lipid interactions, we investigate the SLB system over longer timescales using an AFM. We start by equilibrating the system far above its $T_m$ (at 40 °C), and then rapidly quench it, with a cooling rate of 1 °C/s, to a set temperature $T_c$ where the membrane is still fluid in solution but not necessarily when supported. To choose the $T_c$ we consider the onset of the proximal leaflet transition (27.2 ± 0.6 °C), and set it a few degrees below, either $T_c$ = 25.0 ± 0.1 °C or $T_c$ = 23.0 ± 0.1 °C. We subsequently image the SLBs evolution over a period of 24 h, sealing it inside a chamber to prevent evaporation of the liquid and maintaining $T_c$. From a thermodynamics perspective, the experiment is designed to follow the membrane evolution through an isothermal phase transition at equilibrium. For a mono-component lipid membrane, we observe the expected nucleation and growth of $S_o$ domains until the observed area is fully covered, albeit with a relatively slow kinetic due to the substrate-SLB interactions (suppl Fig. S5). Significantly more complex behavior is visible for our model *E. coli's* binary membrane in Figure 3A, despite the DSC measurements indicating a single, full transition. The early stages of the transition follow the expected classical nucleation and growth behavior, with initially small, randomly distributed $S_o$ domains that grow over time. However, between 1 and 3 h we observe a sudden slowdown in the growth of the $S_o$ domains, coupled with the formation of new thinner domains within the $S_o$ domains. Given their reduced thickness compared to $S_o$, we interpret these as $L_d$. The observed reorganization of the membrane not only stops the initial phase transition, but also transforms the morphology of the domains over 24 h in a manner reminiscent of a spinodal decomposition (102). Experiments with the bilayer always maintained fluid (~4 °C above $T_m$ of the proximal leaflet) did not show any evolution over >3 h (suppl. Fig. S6).
Contacts between bilayers in MLVs could, in principle, allow a similar membrane reorganization. This would affect the membrane internal free energy and hence likely the shape of the DSC profile around the transition peak. This possible molecular reorganization is different in nature than the substrate-induced slowdown of the phase transition kinetics discussed previously, and where only the position of the transition maximum was examined. Here, comparing the overall shape of the cooling and heating phase transition profiles observed by DSC (Fig. 3B, C and suppl. Fig. S7) reveals differences in shape that depend on the scanning rate. A larger difference in shape is observed at the slower scanning rates of 2 °C/min and 5 °C/min, and almost disappears at 10 °C/min (Fig. 3C, see also Fig. S7D). This is consistent with the expected molecular reorganization of the membrane, with an associated timescale is in the order of ~1 min, much faster than observed by AFM for SLBs on mica. Allowing the system to equilibrate for up to 1 h at ~2 °C below their $T_m$ has no impact on the profiles difference if subsequently scanned at 5 °C/min (suppl Fig. S8 and Table S3), suggesting no further membrane evolution on the longer term. Overall, the DSC results are consistent with a relatively fast (~1-2 min), contact-induced rearrangement that occurs when the membrane is below its $T_m$, and can be observed if the cooling/heating rate is sufficiently slow to let the rearrangement take place. However, because DCS results are indirect and here operating at the instrument limit in terms of sensitivity and scan rate (curves quantization in Figs 3B and S7B), we cannot exclude other possible interpretations.

**Surface interactions induce spinodal-like reorganization of the SLBs**

To obtain a more quantitative handle on the observed molecular rearrangements, we conducted a scaling analysis of the system's evolution, focusing on the AFM images (Fig 3A). The underlying hypothesis is the existence of a length scale $L(t) \propto t^\alpha$ with $t$ the time and $\alpha$ characteristic of the domain growth process at play (103, 104). The definition of $L(t)$ is not unique, but it is generally related to the structure factor $S_k$ of the system, defined as an average over $k$ values $S_k(t) = \langle \phi(k,t)\phi(-k,t) \rangle$ with $\phi(k,t)$ the Fourier transform of the AFM image (103, 105). We then calculate $L(t)$ as follows (105):

$$L(t) = 2\pi \frac{\int S_k(t)\, dk}{\int k\, S_k(t)\, dk} \qquad (1)$$

When tracking the evolution of the system supported by an atomically flat mica surface (Fig. 3A), plotting $L(t)$ vs $t$ in a log-log representation reveals three distinct regimes (Fig. 3E). First, a classical diffusion-limited growth takes place over ~45 min, characterized by $\alpha = 1/3$ (106). This is followed by a plateau that coincides with the slowdown of the domain's growth and the onset of formation of the secondary domains. Finally, the third regime appears after ~3 h with a decay in the $S_o$ phase in favor of the $L_d$ phase and is characterized by $\alpha = -1/3$, a value previously observed in spinodal processes (105). Given the spontaneous nature of spinodal processes (104), its formation in the later stages of the experiment can only be explained by a particularly slow kinetics or the late formation of suitable conditions. In fluids, several molecular reorganization mechanisms can coexist during the late stages of phase-separation if diffusivity is low (107, 108). In the case of SLBs, studies have shown that substrate interactions can just moderately reduce lipid diffusivity (at max ~2 times slower than GUVs (109–114)), ruling out diffusion-related kinetic effects. Additionally, the fact that classical nucleation, growth and spinodal processes all appear consecutively in the same system suggests the intervention of a different mechanism. Consistently, repeating the experiment at different quenching temperatures uniformly changes the kinetics of the process through all three regimes, with slower kinetics the further the system is from $T_m$ as seen in Figure 3E.

**Surface induced flip-flop explains the molecular mechanism behind the spinodal reorganization**

Both the DSC results and the AFM experiments are consistent with a molecular reorganization of the bilayer induced by a physical contact with either another membrane or a substrate. In the case of SLBs observed with AFM, this reorganization occurs through spinodal process during the third regime of membrane evolution. We propose that it is induced by substrate-mediated inter-leaflet flip-flop of the lipids (80, 87, 88, 115, 116), resulting in domains with different lipid composition: depleted or enriched in one of the two lipid species initially present evenly across the membrane. While the lateral and rotational diffusion occur rapidly in synthetic and native biological membranes (117, 118), flip-flop process is considerably slower when without the involvement of enzymes (119, 120) because of the high energetic cost associated with the hydrophilic headgroup passing through the hydrophobic interior of the bilayer. Consequently, it can be significantly affected by environmental parameters such as temperature (80), the presence of defects in planar bilayers (89), and of a solid substrate (84, 87, 88). The time scale of the third regime observed here, together with the presence of the substrate, indicate that the surface influences or bias the spontaneous flip-flop process (80, 86, 121).

Directly observing of the resulting lipid species reorganization would require high-resolution chemical mapping and our attempts to label a single specie after the spinodal decomposition proved unsuccessful (see Supporting Information 4.1. for details). Alternatively, techniques such as neutron diffraction (122–124), neutron reflectometry (86) or nano-Raman in solution (125) offer some chemical sensitivity, but they are particularly challenging for achieving single leaflet resolution on unmodified SLBs. We therefore opted for tracking the evolution of membrane's mechanical properties using AFM (Fig. 4). While indirect, the method offers nanometer spatial resolution and the local membrane mechanics depends on its composition (126). Domain-specific evolution can hence provide a signature of the associated molecular rearrangements. To correlate the membrane's mechanics at any given location with its corresponding domain, we used force spectroscopy mapping of evolving SLBs at different times, in a similar fashion as depicted in Figure 3A. Specifically, we calculate the membrane rupture force ($F_R$) and Young's modulus ($Y_M$) from the spectroscopy curves (94, 127) (see *Materials and Methods*). As expected $S_o$ regions appear stiffer and harder to break compared to the $L_d$ phase (127, 128). However, comparing the ratio of the average $F_R$ (Figure 4D) and $Y_M$ values (Figure 4E) obtained over the apparent $L_d$ and $S_o$ domains reveal a significant stiffening of the $S_o$ phase compared to the $L_d$ regions which occurs rapidly after ~3 h, i.e. at the onset of the spinodal process (see also Fig. S9 and Table S4).

Next, we add more evidence to support the proposed surface induced mechanism. In Figure S10 we confirm that the third regime is only observed when at least two different lipids are present on the substrate. In Figure 5A (see also Fig. S11) we show that the supported lipid monolayers on highly oriented pyrolytic graphite (HOPG) with the same binary composition, only undergo the phase transition in the first regime, but with no spinodal process taking place afterwards. Lastly, we repeat the experiments with SLBs on a different substrate to alter the lipid-substrate interactions and hence the presumed bias over the flip-flop (Fig. 5B-C). We selected oxidized PDMS (Fig. 5B, see *Materials and Methods* for preparation details) because its different surface chemistry and increased roughness enables higher lipid mobility and local curvature known to enhance the flip-flop rate (84), but also because the material's properties are tunable (129, 130), allowing its stiffness to be reduced to better match that of biological structures such as the peptidoglycan cell wall ($Y_M$ ~3–50 MPa (131, 132)). While the formation of defect-free membranes and AFM experiments are more challenging on PDMS (130), we observe qualitatively similar behavior as presented in Figure 3A (Fig 5B, see also suppl Fig S12). The $S_o$ domains form too rapidly to allow imaging of the first regime, but topographical features consistent with the spinodal process can already be observed after ~30 min, with the formation of the $L_d$ domains inside the $S_o$ phase. The SLB's poor stability made it impossible to follow the membrane for 24 h and the deduced $L(t)$ values exhibit a larger uncertainty, but the results nonetheless confirm the same phenomenon as on mica, albeit with a faster dynamic. The linear fit yields a growth exponent greater than –1/3, still indicating formation of new $L_d$ domains but deviating from the exponent previously observed on mica (Figure 3E). The fastest substrate-induced evolution was observed for lipid bilayers supported by another bilayer (Fig. 5C), appearing within minutes of quenching the temperature below $T_m$ and yielding $\alpha \approx 0$ in Figure 5D (blue dashed line). The results either indicate that the system has already completed its rearrangement, or that we are simply not able to tracking of $L(t)$ overtime accurately due to the uneven surface of the stacked bilayers (see also suppl. Figure S13 for more details). The rapid molecular evolution here observed is consistent with the DSC measurements on MLVs in Figure 3B-D. The absence of 2D confinements for DSC measurements can further enhance the flip-flop induced rearrangement by allowing fluctuations and curvature also in the third dimension (84).

**Discussion**

Most natural biomembranes are constantly in contact with their surroundings, from the cell wall or the cytoskeleton to compartments within the cell. Here we investigate the impact of such 'physical' contact on the molecular organization and evolution of a model inner membrane of *E. Coli*. Our results indicate that interactions between the membrane and the surface can passively but dramatically affect both the local molecular composition, and the membrane mechanics. The relatively weak and non-specific interaction between lipids and the surface bias the system towards a different equilibrium. To the best of our knowledge, such a complex evolution has never been reported in lipid membranes.

The effect can alter the local molecular diffusion within the membrane, and the substrate-influenced flip-flop provides a mechanism to reorganize the lipid composition and induce leaflet asymmetry in the contact area, with the associated timescale depending on the properties of the contacting surface.

Given the fact that our experiments are conducted *in-vitro* and on model membranes, the prevalence of this effect in natural systems remains to be assessed. However, our results make it possible to identify some of the key thermodynamic and kinetic conditions and characteristics of the process. First, the molecular reorganization takes place if the membrane has one leaflet in the $S_o$ phase, which requires it to be close to its $T_m$ but remain above it in the absence of contacts (here up to ~8 °C). When below the supported proximal leaflet $T_m$, the speed of the spinodal process scales with the quenching temperature (Fig. 3), consistent with the idea of a flip-flop-limited evolution. Second, the effect depends on the balance of lipid-lipid and lipid-substrate interactions (Fig. 5). This implies that both the lipid composition of the membrane and the surface chemistry and topography of the contacting surface determine the extent and the kinetics of the molecular reorganization. Thus, careful control of the phase behavior is needed for studies relying on SLBs since their slow phase kinetics may preclude reaching equilibrium over the course of the experiments, depending on the temperature history of the system and its substrate.

The role of 'contact-induced' molecular reorganization in more realistic biological membranes (133) and *in vivo* will need to be assessed in the future, but the characterized impact of the spinodal process on the biophysical properties of the membrane provided in this work suggests it could be highly relevant for understanding the function of prokaryotic membranes. The change in the local mechanical properties associated with the spinodal decomposition is considerable. Furthermore, a punctual area of contact, e.g. anchoring or support point such as those of FtsZ ring (62, 134, 135) and MreB actin-like filaments (60, 61, 136), can create a local singularity in terms of diffusion or geometry, which in turn may enhance the flip-flopping rate and trigger a rapid local molecular reorganization. Our experiments also highlight the fact membrane-membrane interactions –arguably one of the most common forms of physical contact between biological membranes– can dramatically and rapidly alter the local biophysical properties of the membrane.

**Materials and Methods**

**Lipids and chemicals**

The commercially available lipids, 1-palmitoyl-2-oleoyl-sn-glycero-3-phosphoethanolamine (POPE) and 1-palmitoyl-2-oleoyl-sn-glycero-3-phospho-(1'-rac-glycerol) (POPG), were purchased and dissolved in chloroform from Avanti Polar Lipids (Alabaster, AL).

Salts (all>99% purity) were purchased from Sigma-Aldrich (Dorset, UK) and dissolved in ultrapure water (Merck-Millipore, Watford, UK) to prepare MOPS buffer-based solution with specific ions concentration as follows: 50mM NaCl, 9.5 mM $NH_4Cl$, 0.5 mM $MgCl_2$, 0.3 mM $K_2SO_4$ and 1µM $CaCl_2 \cdot 2H_2O$. The pH was adjusted to 6.5 prior to mixing with lipids.

**Large multi lamellar vesicles preparation**

10 mg of lipids dissolved in chloroform were mixed 4 mL glass vial following the correct molar ratios. To allow chloroform evaporation, the lipids were then pre-dried under a gentle nitrogen flow and finally placed overnight in a vacuum chamber. Lipids were rehydrated with 1 mL of MOPS buffer-based solution obtaining a final concentration of 10 mg/mL and large multilamellar vesicles were obtained through freeze-thawing. Briefly, the lipid solution was heated while sonicating in a bath sonicator and subsequentially, the lipids were frozen by leaving the glass vial for 15 min in a freezer. Six consecutive cycles of heating-freezing were repeated to ensure vesicles formation, as indicated by the lower turbidity of the final solution.

**Differential Scanning Calorimetry (DSC) measurements**

To observe the effects of the scan rate on the thermodynamic properties of lipid vesicles, DSC measurements were performed on a DSC 2500 (TA Instruments, Delaware, USA). 10 µL of LMVs solution at 10 mg/mL were loaded into the calorimeter and the sample was equilibrated before the measurement for 5 min at the starting temperature. DSC runs on vesicles were performed with variating cooling rate (from 0.03 °C/s up to 0.67 °C/min) within a temperature range of 60 °C to -10 °C. DSC runs were repeated 3 times per sample to ensure reproducibility.
To test compositional rearrangement in the vesicle's membrane, a series of DSC cycles were performed with the same DSC 2500 instrument. After loading the LMVs in the same volume and concentration as previously described, the vesicles were equilibrated at 45°C for 5 min. The sample was then cooled down at 16°C with a rate of 0.08 °C/s and equilibrated at this temperature for a variable amount of time. At the end of this isothermal phase, the sample was heated back up at 45°C with a rate of 0.08°C/s and equilibrated at this temperature before the start of the next cycle.

**Small unilamellar vesicles preparation**

Similarly to LMVs preparation, lipids were mixed and dried in a 4 mL glass vial before being rehydrated in the MOPS buffer-based solution. The lipid solution was gently bath sonicated for 15 min at 45°C, until the solution looked opaque and milky, indicating the dissolving of lipids in the solvent. The solution was extruded 31 times using a Mini-Extruder kit (Avanti Polar Lipids) with 2 Whatman 100 nm filter (GE Healthcare Life Sciences, Little Chalfont, UK) to form small unilamellar vesicles (SUVs). Appropriate final SUVs concentration was prepared depending on the surface used for the supported lipid bilayer preparation.

**PDMS surface preparation**

Sylgard 184 silicon elastomer kit was purchased at Dow Corning Corporation (Michigan, USA). Silicon elastomer and curing agent were gently mixed in a 10:1 weight ratio in a glass container. Subsequently, the mixture was degassed in a vacuum chamber for at least 1 h to remove any air bubble in the solution. The PDMS solution was then deposited on top of a cleaved Muscovite mica disk (SPI Supplies, West Chester, PA, USA) and placed inside a WS-650Mz-23NPPB spin-coater (Laurell Technologies Corporation, Landsalle, PA, USA) to allow a homogenous distribution of the PDMS over the surface. Spin-coating was performed for 10 minutes at 1000 RPM at room temperature. Finally, the PDMS-coated disk was placed in an oven at 50°C overnight to ensure the curing of the PDMS layer.

**SLB and monolayer preparation on different surfaces**

The SUVs solution was diluted to an appropriate concentration to ensure full bilayer or monolayer formation depending on the surface used.  To form SLB on a mica surface, 100 µL of 0.2 mg/mL

SUV solution were deposited on a freshly cleaved Muscovite disk (SPI Supplies, West Chester, PA, USA) placed on the AFM stage and let incubating for 30 minutes at 50°C covered with a Petri dish. To form SLB on PDMS, 100 µL of 1 mg/mL SUV solution were deposited on the cured PDMS surface for 30 min at 50°C. Finally, to form lipid monolayer on top of highly ordered pyrolytic graphite, 100 uL of 0.4 mg/mL SUV solution were deposited on freshly cleaved HOPG surface (SPI Supplies, West Chester, PA, USA) for 30 minutes at 50°C. All the samples were subsequently gently rinsed with the solvent to remove any un-broken vesicle, and the temperature was cooled down at 40°C for 15 min before starting the experiment.

**AFM measurements**

Imaging was conducted using a commercial Cypher ES AFM (Oxford Instruments, Santa Barbara, CA, USA), equipped with temperature control. SNL-10 cantilevers (Bruker Scientific instruments, Billerica, MA, USA) with nominal spring constant of 0.35 N/m were used. The tip has a pyramidal shape with a nominal radius ≤ 12 nm at its apex. The AFM imaging was performed in amplitude mode, where the tip was acoustically oscillated at a frequency close to its resonance while fully immerse in the liquid. By adjusting cantilever oscillations, the imaging conditions were kept as soft as possible ensuring neither tip contamination nor damaging of the sample. Force spectroscopy curves and map were conducted by working in contact mode with a SNL-10 cantilever. Force maps were created from 1024 force curves (32 x 32) over a 6.25 µm$^2$ area. Through the acquired force curved, membrane's mechanical features such as $F_R$ and $Y_M$ were calculated based on commonly used AFM force spectroscopy approaches with thin biological systems(127, 128). The Young's modulus, $Y_M$, has been calculated through a corrected version of the well-established Hertzian indentation model for a semi-infinite medium (137, 138) to take into account the finite thickness of the bilayer and the presence of a hard substrate underneath it (139). To limit surface's contribution the indentation depth was kept below 20% of the bilayer's thickness. We note that commonly used assumptions for the model (e.g incompressible membrane) are not necessarily true for biological systems(140, 141). To still allow for quantitative comparison, we used the same cantilever/tip for all the maps. The cantilever was calibrated before and after the experiment. The inverse optical lever sensitivity by recording a force-distance curve on a stiff mica surface and the spring constant was determined through the cantilever's thermal spectrum.

**Data analysis**

DCS results were analyzed with the TRIOS Software, provided with the instrument. The software was used to normalize the raw heat flow data to the sample mass (this "normalized" heat flow just as "heat flow" in this work), correct thermogram baselines, and extract the corresponding thermodynamic parameters. Here, $T_m$ was identified at the highest point of each calorimetric peak, the calorimetric enthalpy change was calculated by integrating the area under the thermal peak, and the on-set temperature was identified as the first portion of the curve deviating from the DSC baseline. When comparing the shape of the cooling and heating DSC profiles (Fig. 3B-C), a baseline was subtracted (linear fit) and the curves then normalized with respect to their respective maximum.
AFM images were flattened to remove background slope and noise, and analyzed using the Gwyddion software(142), an open-source modular program for scanning probe microscopy data visualization and analysis. The structure factor, $S_k$ and subsequently $L(t)$ were calculated from AFM images using custom-written script in Igor Pro (Wavemetrics, Lake Oswego, USA). Prior to analysis, AFM images were flattened by subtracting a fitted 1$^{st}$-order polynomial background from each scan line, followed by a 2D Fast Fourier transform used to derive $S_k$ through radial averaging in $k$ space. To ensure meaningful interpretation, the $k$ values corresponding to features larger than ½ of the AFM images were progressively removed to avoid edge effects. The scaling behavior of $L(t)$ versus $t$, linear fits were performed on log–log plots, where the power-law relationships appear as straight lines. Fitting regions were selected based on both visual inspection—to identify apparent power-law regimes—and a more systematic approach. For the initial regime, a linear regression with a fixed slope of 1/3 was fitted over progressively larger intervals, and the fit was extended until the $R^2$ value dropped below an acceptable threshold ($R^2$ < 0.98). An analogous approach was applied to the final portion of the curve, using a linear regression with a slope of –1/3.

AFM force spectroscopy data were fitted using a custom script written in Igor Pro, allowing for the automated identification of the $F_R$ and $Y_M$ (94).

Graphs were generated using Igor Pro Software (Wavemetrics, Lake Oswego, OR, US) and Python(143).

## Acknowledgments

NT and KV acknowledge funding from the Engineering and Physics Sciences Research Council (EPSSRC grants EP/S023631/1 and EP/S028234/1, respectively). We thank Alexander Morozov and Simun Titmuss for useful discussions, the Biophysical Science Institute at Durham University and Denise Li from the Edinburgh Complex Fluids Partnership for help with the calorimetry measurements

**Figures and Tables**

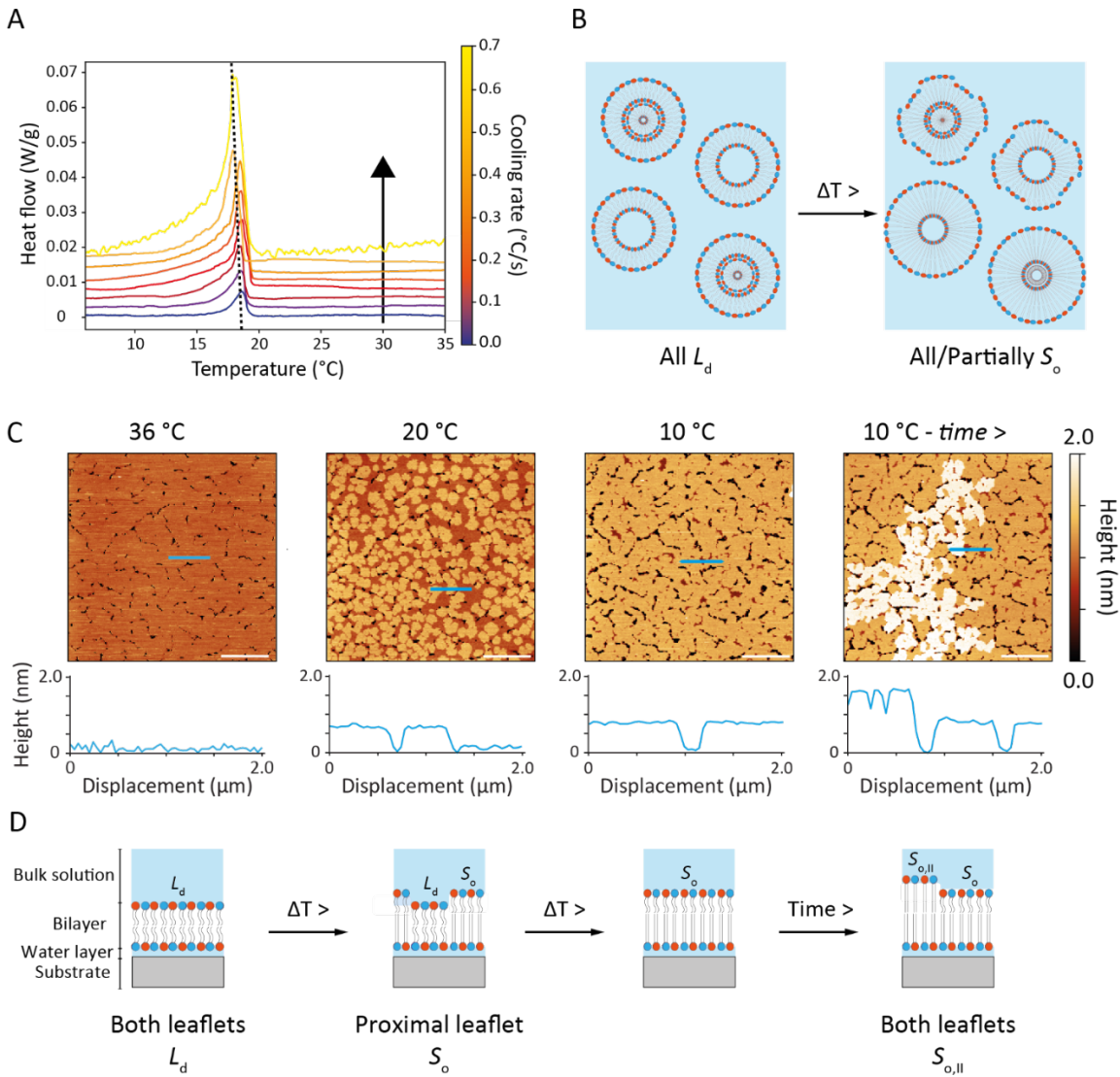

**Figure 1.** Calorimetric analysis of unsupported lipid vesicles and AFM analysis of SLBs. (A) DSC thermographs on lipid vesicles at different cooling rates. The curves have been vertically offset for clarity, evidencing the peak shift to lower temperatures as the rate increases. The derived equilibrium transition temperature is Tm = 20.0 ± 0.2°C (83). A cartoon of the measured LVs is shown in (B), with the sample containing both unilamellar (ULVs) and multilamellar vesicles (MLVs) in solution undergoing the phase transition. In the case of the SLB's, the phase transition is observed by AFM in solution (C): when the lipids arrange into the So phase, the resulting membrane domains appear taller. However, the transition of SLBs is decoupled between the two membrane leaflets resulting in two distinct Tm. Here we consider the Tm of the leaflet in contact with the substrate. The proximal leaflet transitions at higher temperatures due to stabilizing interactions with the substrate and reduced configurational entropy. The transition is fully complete when both leaflets are in So phase, requiring lower temperatures than for the proximal leaflet (here shown at 10° C). The process is illustrated as a cartoon in (D). The scale bar in (C) is 2 µm.

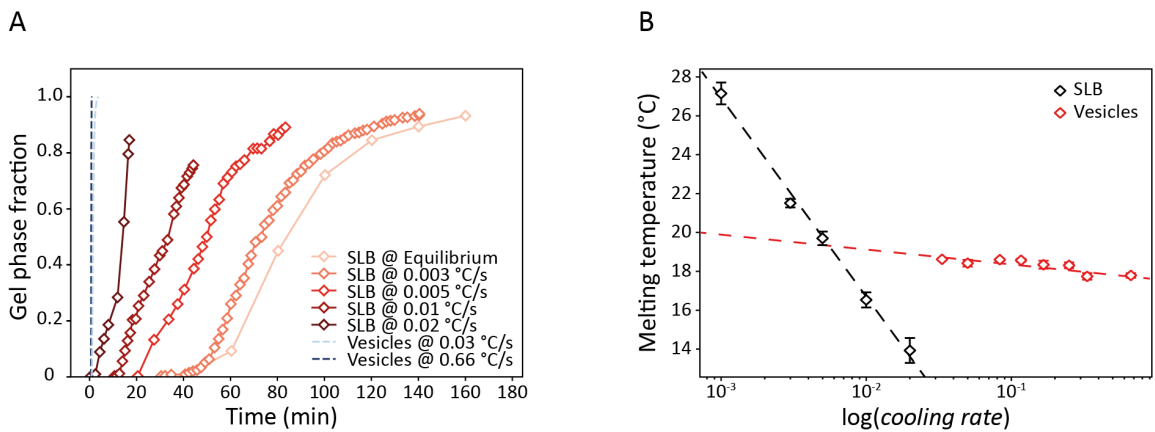

**Figure 2.** Comparison between the effect of the cooling rate on the phase transition of unsupported vesicles and SLBs. (A) Kinetics of the phase transition process for LVs (dashed lines) and the proximal leaflet of SLBs (diamond marked lines) at varying cooling rates. The curves have been horizontally offset for clarity. (B) Semi log plot comparing the Tm variation as a function of cooling rate in SLB and vesicles' systems. The dashes lines represent exponential fits of the experimental data, while non logarithmic graphs are presented in Figure S2B, and C and Table S1, S2. The equilibrium Tm value obtained for LVs coincides with that inferred at 10-3 °C/min, suggesting no significant kinetic effects at that rate. The mean values represented in (B) are given with the respective standard deviations.

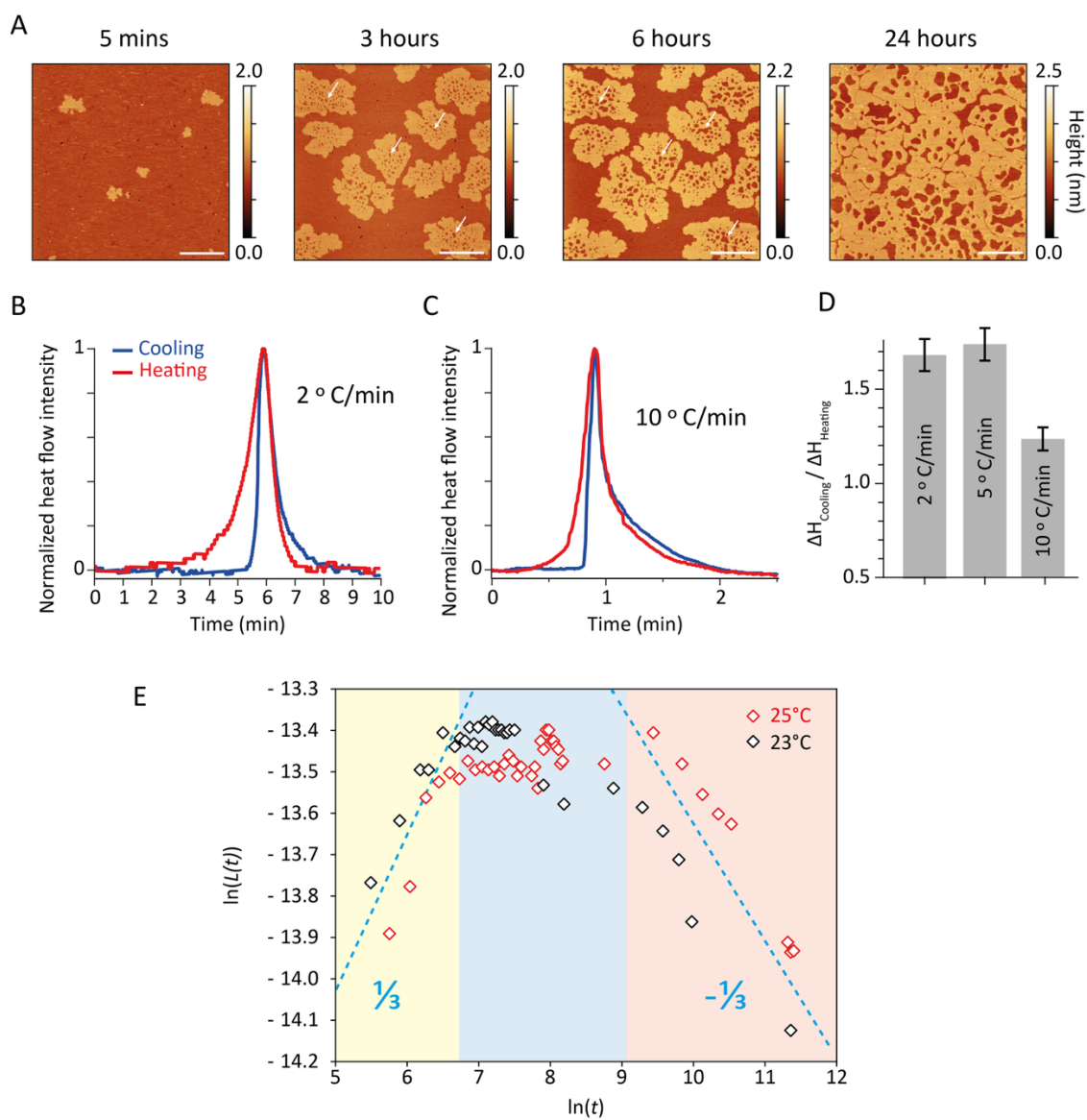

**Figure 3.** Contact-induced reorganization of the membrane. (A) Consecutive AFM topography images of a same SLB region acquired at different time intervals. The temperature is fixed at 25 °C. The scale bars are 2 µm. (B) Comparison between the normalized cooling and heating phase transitions profiles obtained by DSC at 2 °C/min with significant differences between the profiles. (C) The differences visible in (C) largely disappear at 10 °C/min. (D) Enthalpy ratio between the normalized heating and cooling transitions at different DSC scanning rates (see suppl Fig S7 for raw DSC data at 5 °C/min). (E) Temporal evolution of the membrane's structure factor $L(t)$ over 24 h, as derived from AFM images such as presented in (A). $\ln(L(t))$ is plotted vs $\ln(t)$ for two different fixed temperatures. The dashed blue lines represent the stated growth exponent. Shaded colored backgrounds have been used to indicate the boundaries between the three regimes of the process.

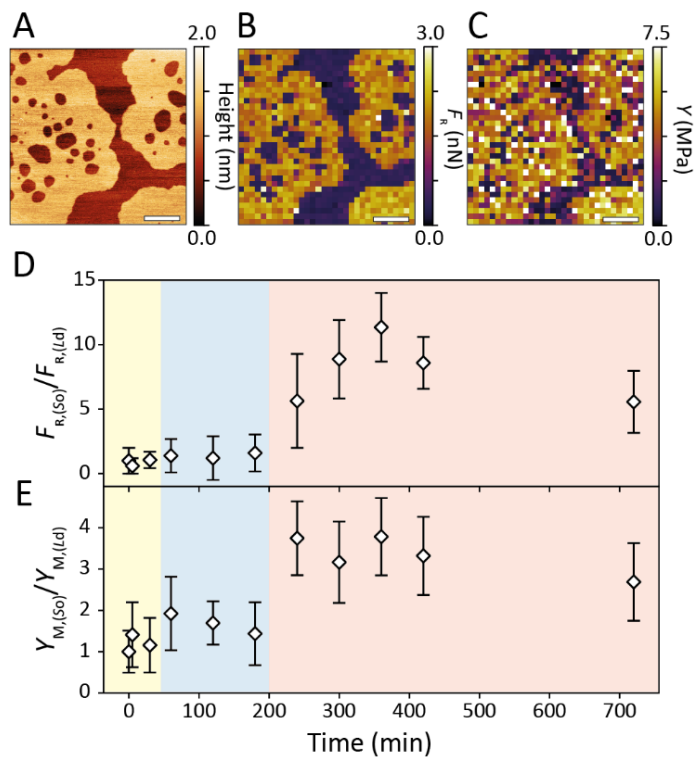

**Figure 4.** Mechanical evolution of the $S_o$ and $L_d$ domains over time. (A) AFM topography image of a binary bilayer on mica 6 h from the start of the experiment. (B-C) AFM force maps displaying rupture force values and young's modulus values respectively. The scale bar is 500 nm. (D) and (E) depict the ratio of gel and liquid $F_R$ and $Y_M$ plotted against time, respectively. The absolute values of $F_R$ and $Y_M$ for both phases can be found in Figure S9 and Table S4. Shaded backgrounds matching Figure 4B have been added to indicate the 3 regimes. In the ratio's plots (D-E), means are given with standard deviations.

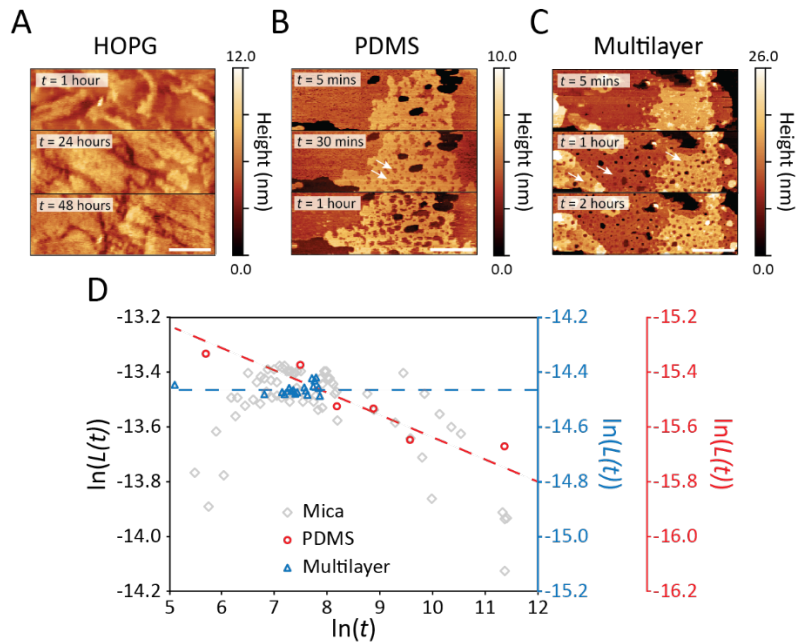

**Figure 5.** Evolution of Binary Lipid Monolayers and Bilayers on Different Substrates Observed via AFM. Upon quenching the temperature below the phase transition point, monolayers deposited on (A) HOPG exhibit slow domain growth, initiated primarily from substrate step edges. These domains eventually coalesce to cover the entire surface, with no evidence of $L_d$ domains forming at any stage. In contrast, rapid phase separation is observed in monolayers on (B) PDMS and in (C) stacked bilayers. On PDMS, $S_o$ domains begin to form within approximately 30 min of cooling, while in the multilayer system, domain formation occurs almost immediately after the temperature drops below the transition point, as indicated by the white arrows. Full AFM images and corresponding height profiles for each system are provided in Figures S11-13. (D) The evolution of $L(t)$ for lipid bilayers on PDMS and in multilayer systems is compared with previously reported data for mica-supported membranes. For the PDMS-supported system, $S_o$ domains formation results in a negative trend similar to that observed on mica, but with a different exponent (red circle marks and red dashed line). For the multilayer system, we obtain α≈0 either because phase separation occurs too quickly or because the presence of defects prevents the formation of a smooth, continuous surface, making it challenging to track $L(t)$ accurately. Scale bars: 200 nm (A), 500 nm (B), and 2 μm (C).

**Supporting Information for**

# Contact-induced molecular reorganization in *E. coli* model lipid membranes

Nicolo Tormena[1], Teuta Pilizota[2,3,*], Kislon Voitchovsky[1,*]

1. Physics Department, Durham University, South Road, Durham DH1 3LE, UK
2. School of Biological Sciences and Centre for Engineering Biology, The University of Edinburgh, Alexander Crum Brown Road, Edinburgh, EH9 3FF, UK
3. Department of Physics, University of Cambridge, JJ Thomson Avenue, Cambridge CB3 0HE

Correspondence: tp579@cam.ac.uk, kislon.voitchovsky@durham.ac.uk

**This file includes:**

- Figures S1 to S13
- Tables S1 to S4
- Supplementary Note 1: Attempt to chemically label the membrane to track its molecular organization
- Supplementary References

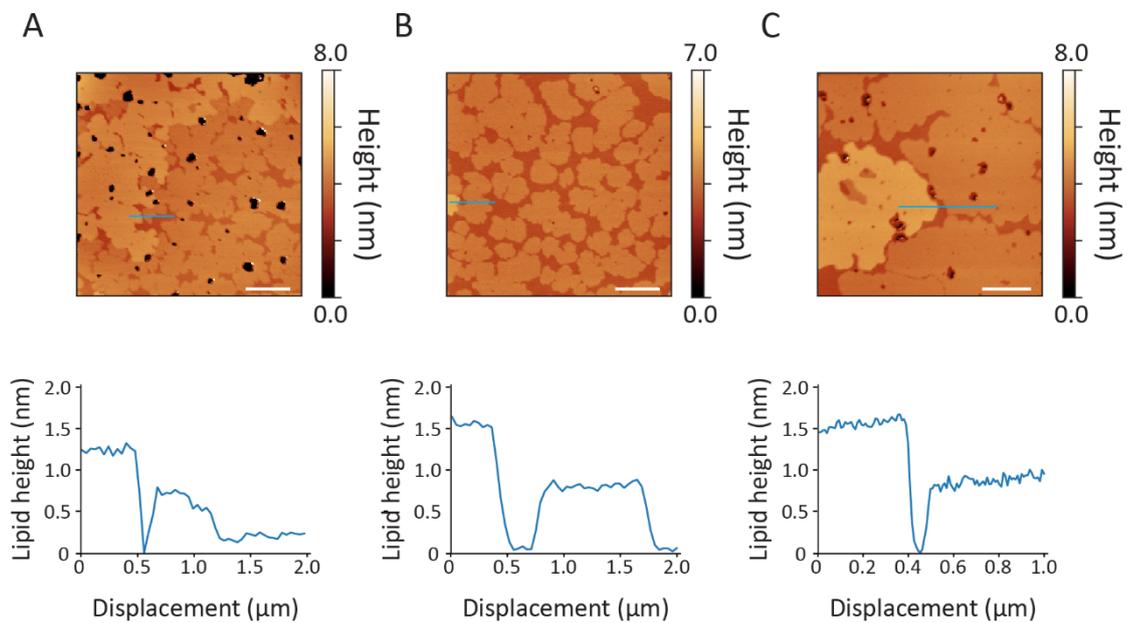

**Figure S1.** AFM images of the second leaflet transition in SLBs at lower temperatures. (A-C) AFM topography examples of the transition of the second leaflet of the binary SLB systems at different cooling rate at approximately 10°C. The transition was driven with a cooling rate of 0.005 °C/s (A) and 0.003 °C/s (B-C). Scale bars are 2 µm in (A-B) and 500 nm in (C). Below each AFM image is the height profile obtained along the blue lines indicated in the image.

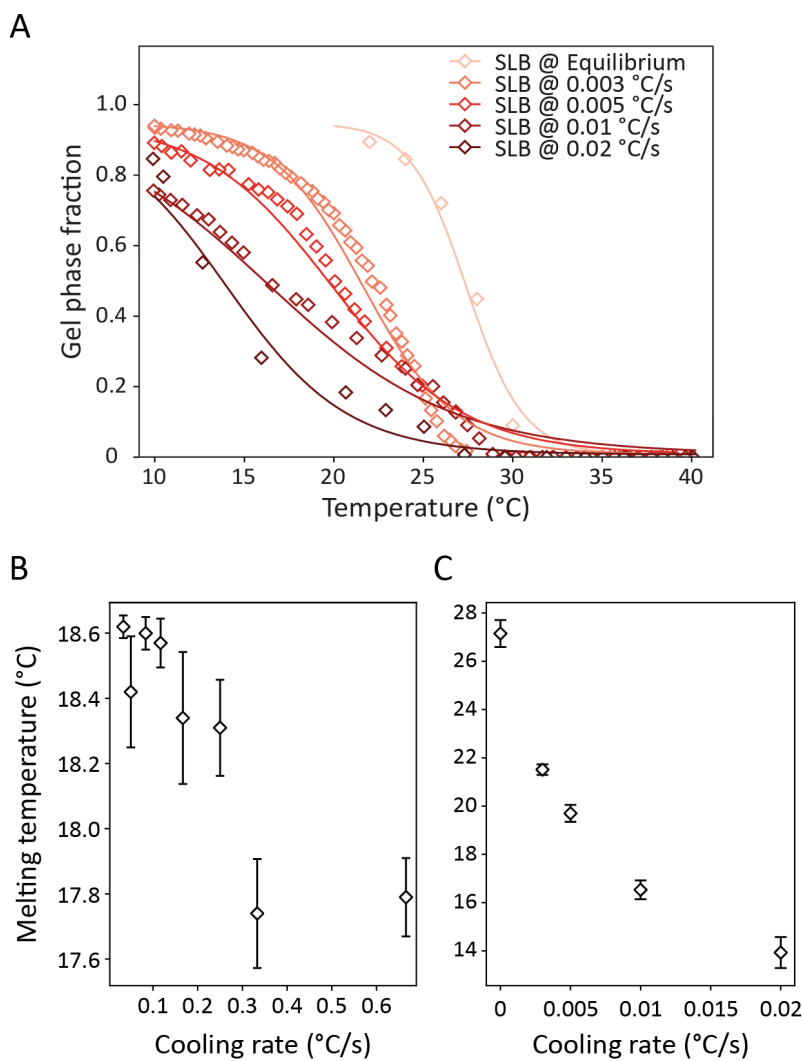

**Figure S2.** Van't Hoff fit for the analysis of the AFM thermographs and comparison of $T_m$ values calculated for both vesicles and SLB systems. (A) Gel phase fraction growth as a function of temperature with Van't Hoff equation fits for SLB performed as in (1):

$$\ln(k) = \frac{\Delta H_{vH}}{R}\left(\frac{1}{T_m} - \frac{1}{T}\right) \qquad (1)$$

where $\Delta H_{vH}$ is the Van't Hoff enthalpy associated with the reaction, and $k$ the equilibrium constant for the $S_o$-to-$L_d$ transition calculated from the gel phase fraction. Experimental data (diamond marks) are displayed with the fitting curves (lines) which have been used to obtain the melting temperature and the Van't Hoff enthalpy associated with the phase transition in each condition. The significant variation of the $T_m$ as a function of the rate applied suggests an out-of-equilibrium condition at faster rates. As mentioned in main text, the here called "equilibrium" condition represents a quasi-static cooling experiment where temperature was decreased every 15 minutes by 2 °C, providing time for the system to equilibrate at each step. Experimental melting temperatures of LVs (B) and SLBs (C) as a function of the applied cooling rate. While the $T_m$ of LVs shows minimal variation—except for a slight shift at cooling rates approaching the instrumental limit—the $T_m$ of SLBs is highly sensitive to cooling rate, exhibiting a total difference of 13 °C.

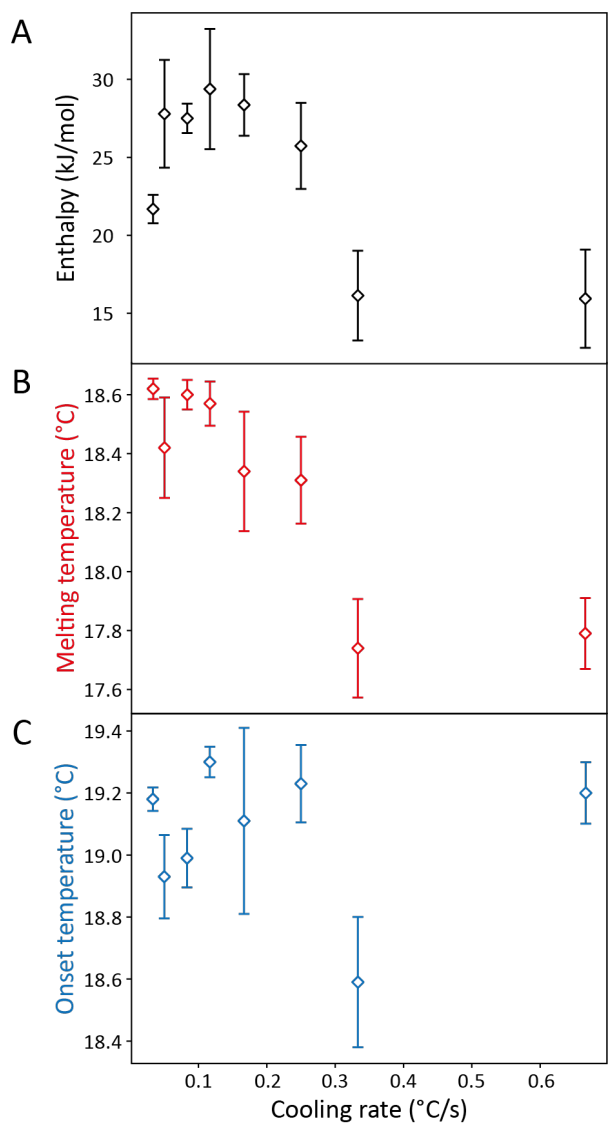

**Figure S3.** Thermodynamic properties of LVs as a function of the cooling rate during DSC experiments. (A) Enthalpy associated with the phase transition, (B) the melting temperature, (C) and the onset temperature of the transition. The two highest rates approach the DSC limit of 0.33 °C/s and 0.67 °C/s. The graphs represent the average with standard deviation for each condition.

| Cooling rate (°C/s) | Melting temperature (°C) | Change of enthalpy (kJ/mol) | Onset temperature (°C) |
|---|---|---|---|
| 0.03 | 18.62 ± 0.03 | 21.68 ± 0.91 | 19.18 ± 0.04 |
| 0.05 | 18.42 ± 0.17 | 27.79 ± 3.46 | 18.93 ± 0.13 |
| 0.08 | 18.60 ± 0.09 | 27.50 ± 0.94 | 18.99 ± 0.09 |
| 0.12 | 18.57 ± 0.05 | 29.38 ± 3.86 | 19.30 ± 0.05 |
| 0.17 | 18.34 ± 0.30 | 28.36 ± 1.98 | 19.11 ± 0.3 |
| 0.25 | 18.31 ± 0.12 | 25.73 ± 2.76 | 19.23 ± 0.12 |
| 0.33 | 17.74 ± 0.21 | 16.13 ± 2.88 | 18.59 ± 0.21 |
| 0.67 | 17.79 ± 0.10 | 15.93 ± 3.15 | 19.2 ± 0.10 |

**Table S1.** The results from the DSC experiments on LVs. Mean with standard deviation is given.

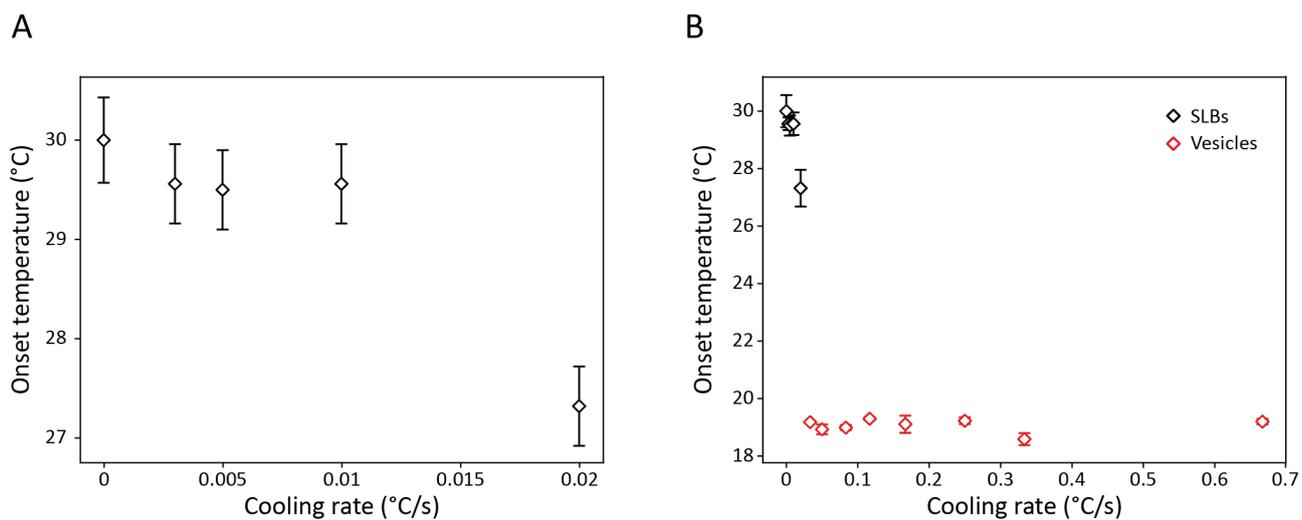

**Figure S4.** (A) Variation of the onset temperature of SLB prepared on mica as a function of the cooling rate, and (B) comparison with previous LVs data (in red). Data shown in the graphs are expressed as mean values, with error bars representing the standard deviations.

| Cooling rate (°C/s) | Melting temperature (°C) | Van't Hoff enthalpy (kJ/mol) | Onset temperature (°C) |
|---|---|---|---|
| Equilibrium (~ 0) | 27.15 ± 0.59 | 448.97 ± 56.7 | 30.00 ± 0.56 |
| 0.003 | 21.51 ± 0.21 | 328.70 ± 10.9 | 29.56 ± 0.22 |
| 0.005 | 19.70 ± 0.20 | 198.80 ± 5.53 | 29.50 ± 0.35 |
| 0.01 | 16.53 ± 0.36 | 123.79 ± 20.9 | 29.56 ± 0.39 |
| 0.02 | 13.93 ± 0.60 | 205.84 ± 14.0 | 27.32 ± 0.64 |

**Table S2.** The melting temperature, Van't Hoff enthalpy and the temperature onset at each cooling rate obtained from the calorimetric AFM experiments on SLBs. Mean values are given with standard deviation.

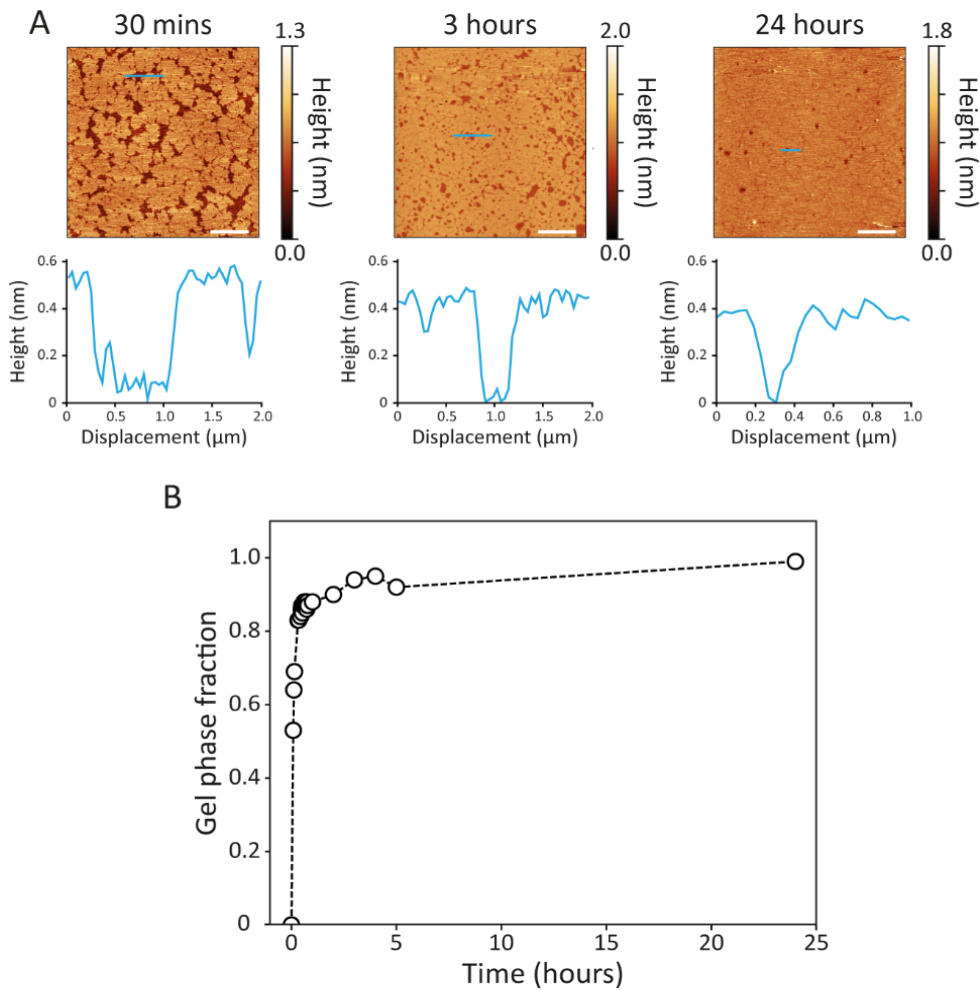

**Figure S5.** POPE SLB undergoes a first order phase transition process within few hours and does not present the same macroscopic morphological features of binary system. (A) Examples of AFM topographical images of POPE SLB at 30°C at different time points after the temperature quench. Each image has been presented with a height profile with scale bar of 2 μm. (B) Gel phase fraction of POPE membrane growth against time after temperature quench. 24 hours from the start approximately 98% of the surface is $S_o$ phase. Both the formation of the SLB and the imaging procedures followed the same protocols described in the *Materials and Methods* section for the binary mixture.

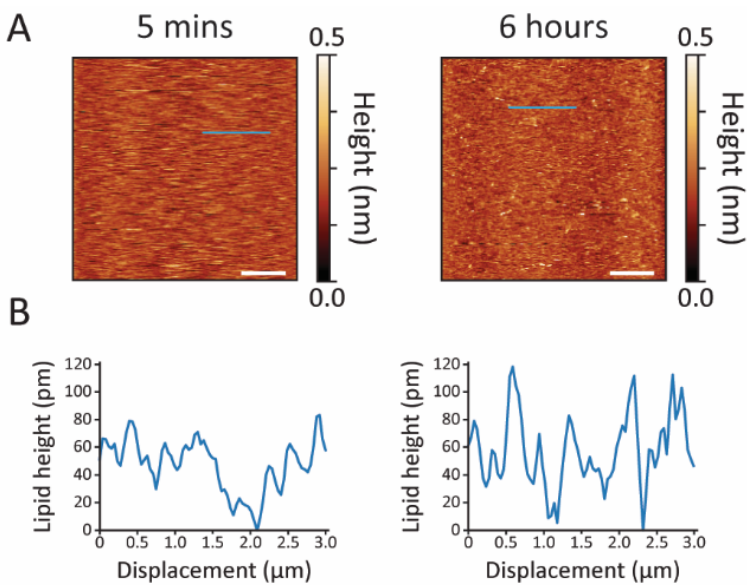

**Figure S6.** Membrane topography does not change as a function of time if the temperature is kept above the mixture's melting point. (A) AFM images of binary lipid mixture on mica at 32°C at different time points throughout the experiment. The scale bar is 2 μm and the temperature has been chosen because it is close to the mixture's $T_m$ but high enough not to trigger the phase transition. (B) Line profiles of the two blue lines traced on the topographical images.

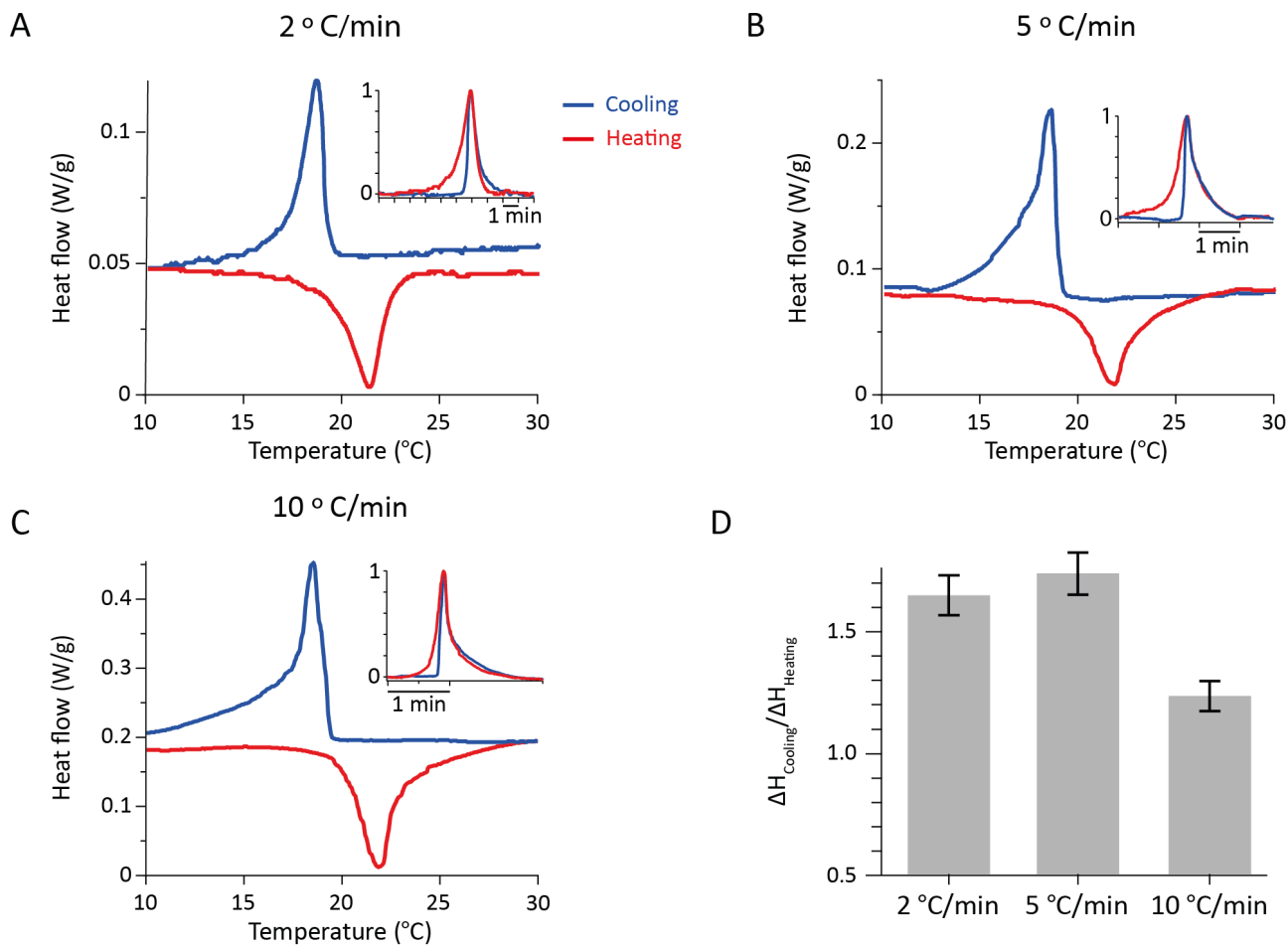

**Figure S7.** Comparison of cooling and heating phase transition curves obtained on LVs and measured by DSC. DSC thermograms of LVs recorded during cooling and heating at (A) 2 °C/min, (B) 5 °C/min and (C) 10 °C/min. Insets display the normalized curves overlaid to highlight differences or similarities in their shapes. The y-axis represents normalized heat flow, while the x-axis corresponds to the time required for the reaction to occur. (D) The ratios of $\Delta H_{cooling}$ to $\Delta H_{heating}$ (with standard deviations) obtained as described in *Materials and Methods* are given for each calorimetric rate.

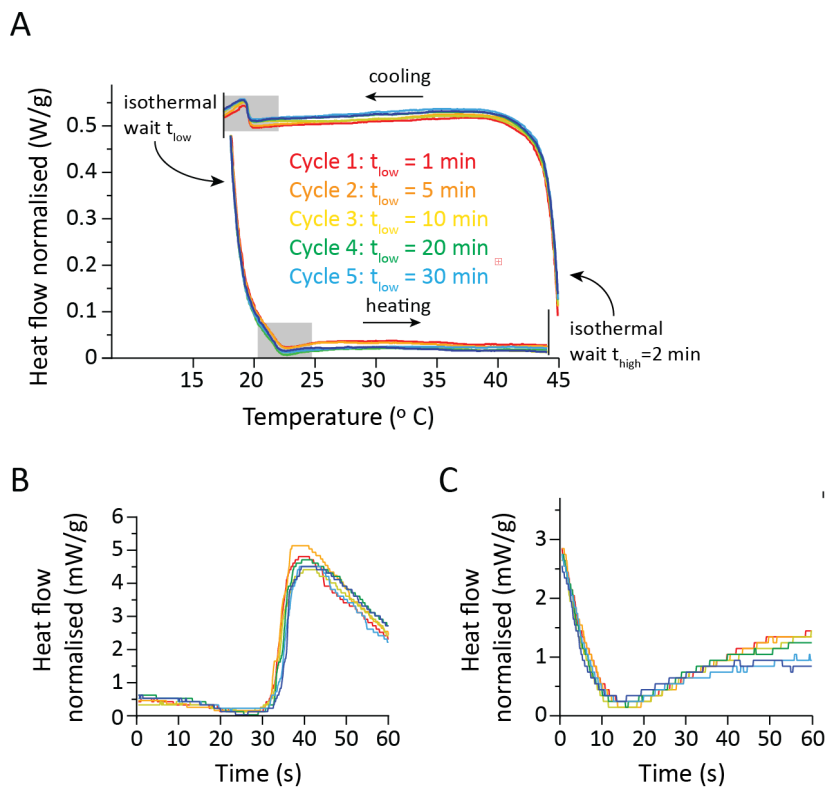

**Figure S8.** DSC cycles with different waiting times at ~2 °C below $T_m$. (A) Normalized heat flow obtained from a sample of LVs first cooled at 5 °C/min (slow enough to allow for membrane equilibration) and kept isothermally for a duration $t_{low}$ at ~2 °C below $T_m$. The sample is then re-heated to control the impact of the isothermal phase on the phase transition properties of the LVs. Six consecutive cycles with different waiting times were conducted, each indicated with a different color. The cooling and heating transitions are highlighted in grey shaded rectangles. The cooling (B) and heating (C) transition peaks are plotted against time. No differences are visible between the different cycles within error, but quantization of the signal (heat flux) indicates measurements are at the limit of the calorimeter.

| Number of cycles | 1st – 1 min isotherm | 2nd – 5 min isotherm | 3rd – 10 min isotherm | 4th – 20 min isotherm | 5th – 30 mins isotherm | 6th – 60 mins isotherm |
|---|---|---|---|---|---|---|
| $T_m$– Cooling (°C) | 18.7 | 18.8 | 18.7 | 18.7 | 18.7 | 18.6 |
| $T_m$– Heating (°C) | 21.9 | 22.0 | 22.0 | 22.0 | 21.8 | 21.8 |

**Table S3.** The melting point of each DSC cycle performed on LVs with varying isothermal time (as defined in Figure S8). Melting points have been evaluated as described in the *Materials and Methods section*. Error bars are not shown, as replicate measurements were not performed owing to the substantial time and liquid nitrogen required for each run.

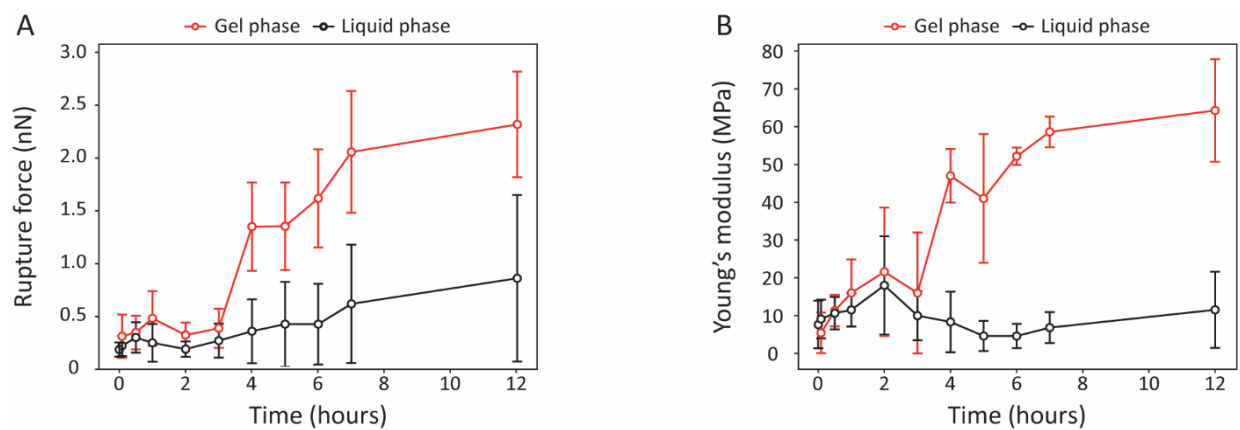

**Figure S9.** Mechanical evolution of SLB's phases as a function of time after temperature quenching. (A) Rupture force values of both $S_o$ phase and $L_d$ phase as a function of time. (B) Young's modulus values of both $S_o$ phase and $L_d$ phase as a function of time. For each condition, a force map consisting of 32 × 32 force curves was acquired. The imaged area typically contained approximately equal fractions of $S_o$ and $L_d$ domains, enabling statistical averaging of mechanical properties for each phase separately. Mean values and standard deviations here plotted were calculated from the respective ~50% of data points corresponding to each phase.

| Time (min) | Rupture force (nN) | | Young's modulus (MPa) | |
|---|---|---|---|---|
| | $S_o$ phase | $L_d$ phase | $S_o$ phase | $L_d$ phase |
| 5 | 0.31 ± 0.21 | 0.22 ± 0.10 | 5.45 ± 5.36 | 9.12 ± 5.13 |
| 30 | 0.35 ± 0.16 | 0.30 ± 0.14 | 11.32 ± 4.19 | 10.65 ± 4.3 |
| 60 | 0.48 ± 0.26 | 0.25 ± 0.18 | 16.01 ± 8.87 | 11.52 ± 4.41 |
| 120 | 0.32 ± 0.12 | 0.19 ± 0.07 | 21.60 ± 17.00 | 18.00 ± 13.00 |
| 180 | 0.39 ± 0.19 | 0.27 ± 0.16 | 16.00 ± 15.36 | 10.00 ± 6.50 |
| 240 | 1.35 ± 0.42 | 0.36 ± 0.30 | 46.99 ± 7.09 | 8.34 ± 7.94 |
| 300 | 1.35 ± 0.41 | 0.43 ± 0.40 | 41.00 ± 17.00 | 4.62 ± 3.79 |
| 360 | 1.62 ± 0.47 | 0.43 ± 0.38 | 52.16 ± 2.28 | 4.60 ± 3.22 |
| 420 | 2.06 ± 0.58 | 0.62 ± 0.56 | 58.61 ± 4.05 | 6.83 ± 4.10 |
| 720 | 2.32 ± 0.50 | 0.86 ± 0.79 | 64.27 ± 13.55 | 11.55 ± 10.05 |

**Table S4.** Values of rupture force and Young's modulus obtained from force spectroscopy measurements on the different phases of SLBs at different time points. Means and standard deviations are given.

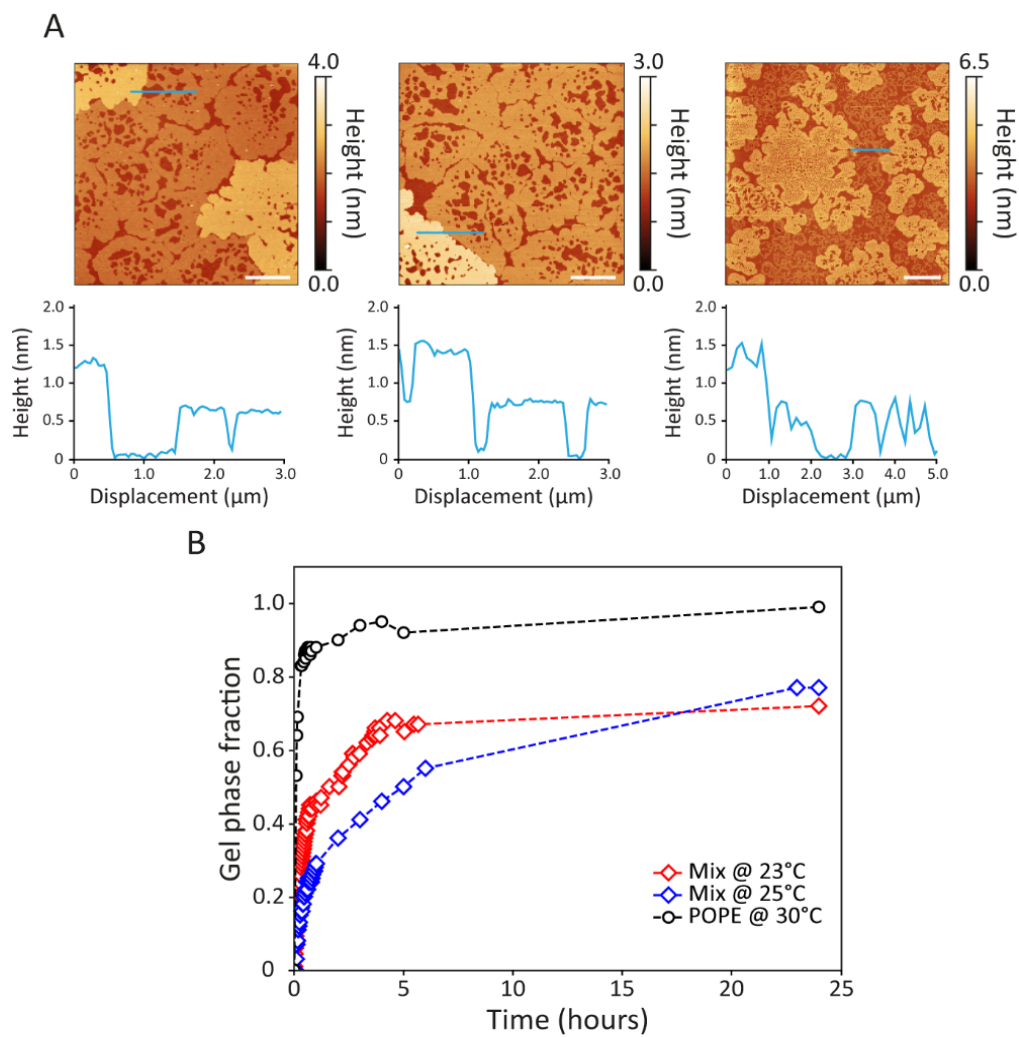

**Figure S10.** (A) Examples of AFM topographical images of bicomponent SLBs at 25°C (left and middle) and 23°C (right) after 24 h from the start of the experiment. Each image has been presented with a height profile underneath it, showing lipid domain's height and the scale bar of the image is 2 μm (left and middle) and 5 μm (right). (B) Gel phase fraction plotted against time after quenching the temperature below the melting point. The two-component model membrane quenched to different temperature is given in red and blue and black gives the pure POPE SLB. The final $S_o$ phase coverage is approximately 75% at both 23°C and 25°C for the mix and 99% for POPE (also shown in Figure S5).

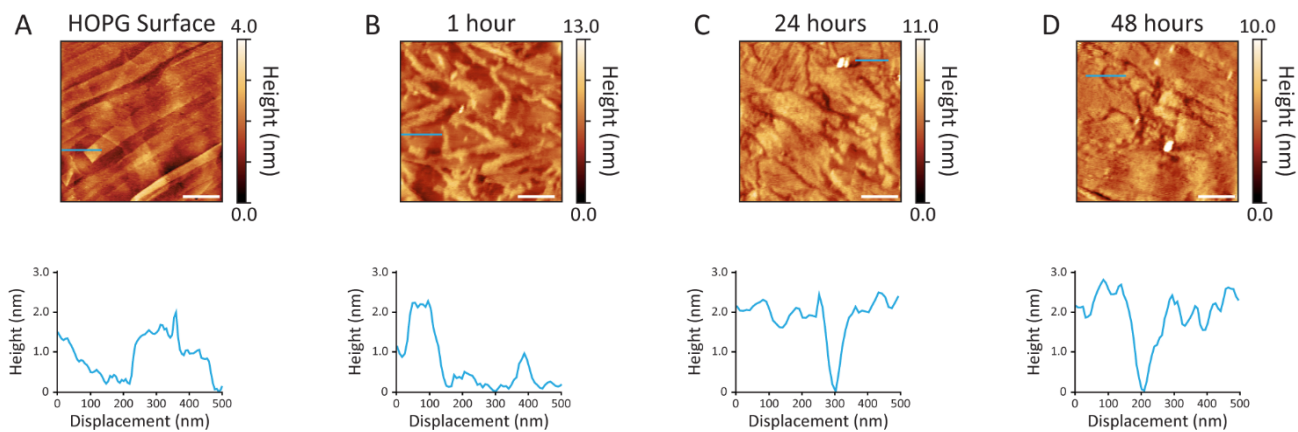

**Figure S11.** Phase transition on binary monolayer prepared on HOPG surface. (A) AFM topography images of HOPG surface without and (B-D) with the lipid monolayer undergoing phase transition over time. Line profiles in blue are presented below each AFM image. The scale bars are 1 μm in (A) and 200 nm in (B-D). Bare HOPG surface is microscopically flat with various layers of sub nanometer steps.

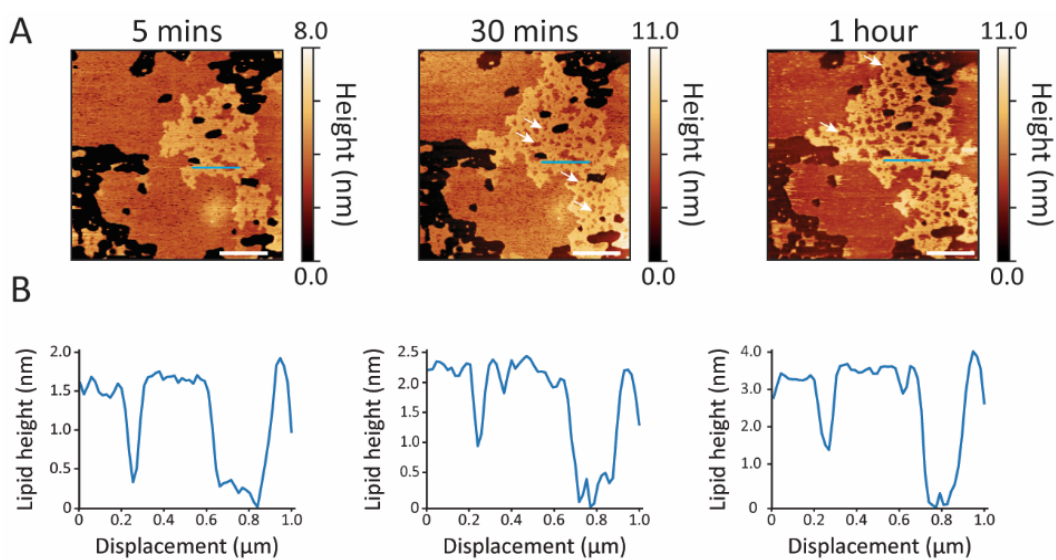

**Figure S12.** Phase transition of binary *E. coli* model membrane SLBs prepared on PDMS surface. (A) AFM topography images of $S_o$ domain growth over time (up to 24h time scales were not accessible as the SLBs are not stable on PDMS surface for that long). White arrows (middle and right images) point at the $L_d$ domains forming within the $S_o$ phase. The scale bar is 500 nm. (B) Height profiles along the blue lines in (A) are given.

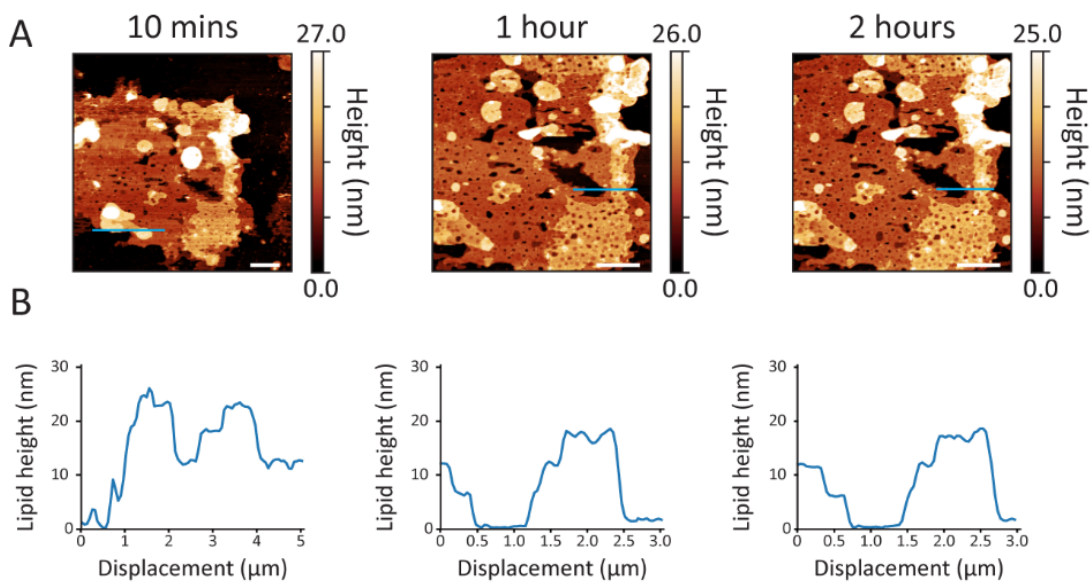

**Figure S13.** Phase transition phenomenon on binary multilayers prepared on PDMS surface with high SUVs concentration. (A) AFM topography images of $S_o$ domain growth over time. The scale bar is 2 µm. (B) Respective height profiles are given below each image.

**Supplementary Note 1: Attempt to chemically label the membrane to track its molecular organization**

Testing the driving force behind the morphological evolution of our SLB systems directly is not trivial, as this would require tracking the movement of single molecules within the bilayer. To observe this hypothesized compositional variation, we attempted to singularly track one of the two lipid species over the different phase transition stages. The first approach involved incubating the membrane, at different time intervals, using cinnamycin, which is an antibacterial peptide able to recognize PE headgroup (2, 3). Although cinnamycin adsorption of the bilayer was observable with AFM, the marker did not spread homogenously across the surface even at high temperature where the mixture was perfectly mixed. Cinnamycin tended to accumulate mainly along bilayer defects, likely due to interactions between it and the underlying surface. As an alternative, we employed cesium ions (positively charged), which we hypothesized would preferentially interact with PG headgroups (negatively charged) due to the favorable electrostatic interactions. Given cesium's large radius, we expected to observe some topographical variations while imaging the sample with AFM. To do that, we worked with a ratio between the set-point amplitude and the free oscillation amplitude greater than 0.6, generally used to perform gentle and high-resolution imaging on AFM (4). Unfortunately, under these conditions, the height and phase images showed no resolvable differences or contrast that could have been attributed to the cesium interaction. he absence of such features is likely due to the weak electrostatic interactions between cesium ions and the membrane, which allow the ions to be displaced by vertical and lateral perturbations from the AFM tip. Changing the solution's pH would have increased the strength of these interactions but inevitably modifying the phase transition properties. Other possible way forward could have included the use of fluorescently labelled lipids. Because this would significantly change the system, in particular the flip-flop rate in the SLBs, we did not attempt it.

**Supplementary References**